\documentclass[a4paper,12pt]{article}

\title{On the exactness of the Semi-Classical Approximation for Non-Relativistic One Dimensional Propagators}
\author{\.{I}brahim Semiz and Koray D\"uzta\c{s} \\
Department of Physics, Bo\~gazi\c{c}i University, Bebek, \.{I}stanbul, TURKEY}
\date{}

\begin{document}

\maketitle

\begin{abstract}
For one dimensional non-relativistic quantum mechanical problems, we investigate the conditions for all the position dependence of the propagator to be in its phase, that is, the semi-classical approximation to be exact. For velocity independent potentials we find that:

(i) the potential must be quadratic in space, but can have arbitrary time dependence.

(ii) the phase may be made proportional to the classical action, and the magnitude (``fluctuation factor'') can also be found from the classical solution.

(iii) for the driven harmonic oscillator the fluctuation factor is independent of the driving term.
\end{abstract}

\section{Exactness of the Semi-Classical Approximation} 
In non-relativistic quantum mechanics one mathematical object of interest is the propagator. It is defined by
\begin{equation} 
\psi(x,t)= \int K(x,t;x_{0},t_{0}) \psi(x_{0},t_{0}) dx_{0} \label{PropDef}
\end{equation} 
and satisfies
\begin{equation} 
\left( H -i\hbar \partial_t \right)K(x ,t;x_0,t_0)=-i\hbar\delta (x-x_0)\delta (t-t_0) . \label{schro1}
\end{equation} 
It can be in principle directly calculated by the path integral technique \cite{feynman}:

\begin{equation}
K(x,t;x_0,t_0)=\int {\cal D}x\;e^{i S/\hbar} \label{pathint} .
\end{equation}

As is proven in any treatment of path integrals (e.g.  \cite{feynmanhibbs,schulman,kleinert1}), 
for some problems such as the free particle and the harmonic oscillator
the propagator takes the form
\begin{equation}
K_{\rm sc}(x,t;x_0, t_0)=f(t-t_0)e^{i S_{\scriptsize{\mbox{cl}}}(x,t;x_0,t_0)/\hbar}
\label{orig0}
\end{equation}
where $S_{\mbox{cl}}$ is the action of the classical path. This is sometimes called the semi-classical approximation since the contribution to the propagator (\ref{orig0}) seems to be from the classical path only, even though the integral (\ref{pathint}) was over all possible paths. 

It was claimed at least once \cite{cb} that {\em all} path integrals reduce to classical paths. The particular argument of  \cite{cb} was shown to be incorrect  \cite{berrymount}, but for which problems {\em is} the semi-classical approximation exact? This question is discussed and left open in \cite{schulman}, where counterexamples are given to argue against universal exactness.

Thus we would like to investigate under what conditions (\ref{orig0}) remains valid. We will assume that the Hamiltonian is of the form
\begin{equation}
H=\frac{p^2}{2m}+V(x,t) \label{ham}
\end{equation}
i.e. we limit ourselves to the case of a potential independent of velocity, but which may depend on time. We consider  $t>t_0$, for which the right-hand-side of (\ref{schro1}) vanishes. Also using (\ref{orig0}) with $S_{\rm cl} \rightarrow S$ and (\ref{ham}), the general equation (\ref{schro1}) reduces in our case to
\begin{equation}
-i\hbar\partial_t \left[ f(t) e^{i S(x,t;x_0,t_0)/\hbar}\right]-
f(t)\frac{\hbar^2}{2m}\partial_x^2
\left[e^{i S(x,t;x_0,t_0)/\hbar}\right]+V(x,t)f(t)e^{i S(x,t;x_0,t_0)/\hbar}=0
\label{orig1}
\end{equation}
For the time being we will take $S$ to be any function of $x$ and $t$, with $x_0$ and $t_0$ as parameters. Equation (\ref{orig1}) gives

\begin{equation}
-i\hbar \frac{\dot{f}(t)}{f(t)}+\frac{\partial S}{\partial t}
+V(x,t) - \frac{i\hbar}{2m}\frac{\partial^2 S}{\partial x^2
}+\frac{1}{2m}\left( \frac{\partial S }{\partial x} \right)^2=0
\label{orig3}
\end{equation}
Since $S$ and $V(x,t)$ are real, the real part of (\ref{orig3}) is
\begin{equation}
\frac{\partial  S}{\partial t}+V(x,t)+\frac{1}{2m} \left(
\frac{\partial S }{\partial x} \right)^2 +\mbox{Re} \left[
-i\hbar\frac{\dot{f}(t)}{f(t)} \right]=0\label{orig4}
\end{equation}
and the imaginary part is:
\begin{equation}
\frac{-\hbar}{2m}\frac{\partial^2 S }{\partial x^2}
+\mbox{Im}\left[ -i\hbar\frac{\dot{f}(t)}{f(t)} \right]=0
\label{orig5}
\end{equation}
from which it follows that $\partial^2 S /\partial x^2 $ is a
function of $t$ only:
\begin{equation}
\frac{\partial^2 S }{\partial x^2}=2F(t) \Rightarrow
S(x,t)=F(t)x^2+G(t)x+J(t) \label{s2f}
\end{equation}
Putting this expression for $S$ and its derivatives into
(\ref{orig4}) and rearranging, we get

\begin{equation}
\left( \dot{F} +\frac{2F^2}{m} \right) x^2+\left(
\dot{G}+\frac{2FG}{m}\right) x+\left( \dot{J}+
\frac{G^2}{2m}+I\right) +V(x,t)=0 \label{orig6}
\end{equation}
where 
\begin{equation}
 I(t)=\mbox{Re} \left[ -i\hbar \frac{\dot{f}}{f} \right] 
\end{equation}
The expressions in parantheses have no $x$ dependence. Therefore
the potential can be written as
\begin{equation}
V(x,t)=F_1(t) x^2+G_1(t) x+J_1(t) \label{orig7}
\end{equation}
So (\ref{orig0}) is satisfied for velocity independent, quadratic
potentials and a general function $S$ in the exponential.

To find the relation of $S$ to the classical path, we must find the classical solution. For this we impose
\begin{equation}
 m\frac{d\dot{x}}{dt}=-\frac{\partial V}{\partial x}=2\dot{F}x+\dot{G}+\frac{2F}{m}(2Fx+G)
\end{equation}
The left hand side is a total derivative therefore the right hand side must also be a total derivative. This gives

\begin{equation}
\frac{2Fx+G}{m}=\dot{x} \label{classsolu}
\end{equation}
In other words we have reduced the problem of the solution of the equation of motion for a given potential of the form (\ref{orig7}) to the solution of a series of first order differential equations: First from $F_1$ and $G_1$ one can calculate $F$ and $G$ via the correspondence of (\ref{orig6}) and (\ref{orig7}), then $x(t)$ can be found via (\ref{classsolu}). These equations are \underline{not} the corresponding Hamilton's equations.

Let us now calculate the Lagrangian as a function of $x$ and $t$:
\begin{eqnarray}
L&=&\frac{m}{2}\dot{x}^2 -V(x,t) \nonumber \\
 &=&\left(\frac{4F^2}{m}+\dot{F} \right)x^2+\left(\dot{G}+ \frac{4FG}{m}\right) x+\frac{G^2}{m}+\dot{J}+I
\label{lagran1}
\end{eqnarray}

On the other hand let us consider the total time derivative of $S(x,t)$:
\begin{eqnarray}
\frac{dS}{dt}&=&\frac{\partial S}{\partial t}+\frac{\partial S}{\partial x}\frac{dx}{dt} \nonumber \\
&=& \dot{F}x^2+\dot{G}x+\dot{H}+(2Fx+G)\dot{x} \nonumber \\
&=&\left(\frac{4F^2}{m}+\dot{F} \right)x^2+\left(\dot{G}+ \frac{4FG}{m}\right) x+\frac{G^2}{m}+\dot{J}
\label{action1}
\end{eqnarray}
where in both (\ref{lagran1}) and (\ref{action1}), we used (\ref{classsolu}) to express $\dot{x}$ in terms of $x$ and $t$.
The expressions for $L$ and $\frac{dS}{dt}$ agree for $I(t)=0$. Therefore we have shown that $S$ is the classical action $S_{\scriptsize{\mbox{cl}}}$ if
\begin{equation}
\mbox{Re}\left( -i\hbar\frac{\dot{f}}{f}\right)=0
\label{condi1}
\end{equation}
i.e. $\dot{f}/{f}$ is real.

To summarize, the propagator can be written in the form
(\ref{orig0}) \emph{only} for quadratic potentials given in (\ref{orig7}). Then the classical action be written in the form (\ref{s2f}), and
the functions that constitute the potential and the classical
action are related by
\begin{equation}
F_1=-\left(
 \dot{F}+\frac{2F^2}{m}\right)\mbox{;}\; G_1=-\left(
 \dot{G}+\frac{2FG}{m}\right) \mbox{;}\; J_1=-\left( \dot{J}+\frac{G^2}{2m} \right) \label{condi2}
\end{equation}

The fluctuation factor $f$ can also be found from the classical solution using
\begin{equation}
\frac{-F}{m}-\frac{\dot{f}}{f} =0 \label{fF}
\end{equation}
 
The integrations for finding $F$, $G$, $J$ from $F_1$, $G_1$, $J_1$ will bring integration constants which must be expressed in terms of $x_0$ and $t_0$, since these are the parameters of $S_{\mbox{cl}}(x,t)$ and don't appear in $V(x,t)$.

To identify these constants we have two conditions at our disposal: First, when solving the classical equation of motion, at the initial time and position, the initial velocity $\dot{x}_0$ must be freely specifiable. Therefore (\ref{classsolu}) must give an undefined result for $\dot{x}$ at $x_0$, $t_0$. The second condition is that $S(x_0,t_0)$ must be zero. 

Equation (\ref{fF}) also means that $f$ is a real function times a complex constant. The constant can be found by considering

\begin{equation}
\lim_{ t\to t_0}K(x,t;x_0,t_0)=\delta(x-x_0) \label{condi3}.
\end{equation}

\section{Applications}

\subsection{Free Particle}
For a free particle $F_1=G_1=J_1=0$, which yields
\[ F=\frac{m}{2(t+C_1)} \mbox{ ; } G=\frac{C_2}{t+C_1} \mbox{ ; }J=C_3+\frac{C_2^2}{(t+C_1)2m}\]
For the classical solution we get
\begin{equation}
\dot{x}=\frac{2Fx+G}{m}=\frac{x+C_2 /m}{t+C_1} \label{inci}
\end{equation}
From the free specifiability of $\dot{x}_0$ we find
\[
x_0+C_2 /m=0 \mbox{ ; } t_0+C_1=0
\]
On the other hand, $S(x_0,t_0)$ condition gives
\[
S(x_0,t_0)=\frac{(mx_0+C_2)^2}{2m(t_0+C_1)}+C_3=\frac{\dot{x}_0}{2m} (mx_0+C_2)+C_3=0
\]which means that $C_3=0$.

Putting $C_1$, $C_2$, $C_3$, the classical action can be rewritten as 

\[
S_{\mbox{cl}}=\frac{m}{2(t-t_0)}(x^2-2xx_0+x_0^2) \]
Applying (\ref{fF}), we find the fluctuation factor $f$
\[ \frac{-1}{2t} -\frac{\dot{f}}{f}=0 \Rightarrow \;
f=C_4/\sqrt{(t-t_0)} 
\] 
Finally the requirement (\ref{condi3}) fixes $C_4$ and

\begin{equation}
K_{\mbox{free}}(x,t;x_0,t_0)=\frac{m}{\sqrt{2\pi i\hbar (t-t_0)}}
e^{\frac{i}{\hbar} \frac{m}{2t}(x-x_0)^2  }
\end{equation}
Incidentally, (\ref{inci}) can be solved to give
\[
x_{\mbox{\scriptsize{cl}}}=x_0+B(t-t_o)
\]

\subsection{The Simple Harmonic Oscillator}
In this case $F_1=mw^2/2$ and $G_1=0$, $J_1=0$. This yields
\[
F=\frac{mw}{2}\tan [w(C_1-t)]\mbox{ ; } G=\frac{C_2}{\cos [w(C_1-t)]} \mbox{ ; }J=C_3+\frac{C_2^2}{2mw}\tan[w(C_1-t)]
\]
Applying (\ref{classsolu}) we get 
\begin{eqnarray}
\dot{x} &=& w\tan[w(C_1-t)]x+\frac{C_2}{m\cos [w(C_1-t)]} \nonumber \\
&=& \frac{mw\sin [w(C_1-t)]x+C_2}{m\cos [w(C_1-t)]}
\end{eqnarray}
Free specifiability of $\dot{x}_0$ gives $w(C_1-t_0)=\pi/2$ and $mwx_0+C_2=0$. The $S(x_0,t_0)=0$ condition gives
\[
S(x_0,t_0)=\frac{x_0mw+C_2}{2w}\dot{x}_0+C_3=0
\]
which again makes $C_3=0$. So $S_{\mbox{cl}}$ becomes
\[
S_{\mbox{cl}}(x,t)=\frac{mw}{2}\cot [w(t-t_0)]x^2-\frac{mwx_0}{\sin [w(t-t_0)]}x+ \frac{mwx_0^2}{2}\cot [w(t-t_0)]
\]
The fluctuation factor $f$ can be calculated as
\[f=C_4/\sqrt{\sin [w(t-t_0)]} \] 
so that the propagator becomes

\begin{equation}
 K(x,t;x_0,t_0)=\sqrt{\frac{mw}{2\pi i\hbar\sin [w(t-t_0)]}}\exp\left\{
\frac{i}{\hbar} \left[ \frac{mw}{2}\left( (x^2+x_0^2)\cot [w(t-t_0)]
-\frac{2xx_0}{\sin [w(t-t_0)]} \right) \right] \right\}
\end{equation}

The classical solution is
\[
x_{\mbox{\scriptsize{cl}}}=x_0 \cos [w(t-t_0)] +\frac{C_5}{m}\sin[w(t-t_0)]
\]

\subsection{The Driven Harmonic Oscillator}
The function $G_1(t)$ in the potential (\ref{orig7}) corresponds to a driving term. If it is non-zero, of course the classical solution and the classical action will change. But the fluctuation factor $f$ depends only on $F$. 

Therefore, the fluctuation factor of a driven harmonic oscillator is the same as that of the undriven one, independent of the time dependence of the driving function, even if the frequency is also time dependent. This verifies and generalizes the result of section 2.7 of \cite{ingold}.

\section*{Acknowledgements}

We would like to thank \.{I}.H. Duru for helpful comments and suggestions. \.{I}.S. acknowledges partial support by grant 06B303 by Bo\~gazi\c{c}i University Research fund.


\begin{thebibliography}{10}
\bibitem{feynman} Feynman, R. P., ``Space Time Approach to Non-Relativistic Quantum
Mechanics'', \emph{Review of Modern Physics}, Vol. 20, pp. 367-387, 1948.
\bibitem{feynmanhibbs} Feynman R. P. and Hibbs R. A. 1965 {\it Quantum Mechanics and Path Integrals} (New York:  McGraw-Hill)
\bibitem{schulman} Schulman L. S. 1981 {\it Techniques and Applications of Path Integration} (New York:  Wiley)
\bibitem{kleinert1} Kleinert, H., \emph{Path Integrals in
Quantum Mechanics, Statistics, Polymer Physics, and Financial
Markets}, World Scientific Publishing Co. Pte. Ltd, Singapore, 2004.
\bibitem{cb} Clutton-Brock M 1965 {\it Proc. Camb. Phil. Soc.}  {\bf 61} 201 
\bibitem{berrymount} Berry M V and Mount K E 1972 {\it Rep. Prog. Phys.}  {\bf 35} 315
\bibitem{ingold}Ingold G L 2002 Path Integrals and Their Applications to Dissipative Quantum Systems  {\it Coherent Evolution in Noisy Environments (Lecture Notes in Physics} vol 611) ed A Buchleitner and K Hornberger (New York: Springer) ({\it Preprint quant-ph/02008026})

\end{thebibliography}
\end{document}